# Photoluminescence enhancement at the vertical van der Waals semiconductor-metal heterostructures

Hafiz Muhammad Shakir[1*], Abdulsalam Aji Suleiman[2*], Kübra Nur Kalkan[2], Amir Parsi[2], Uğur Başçı[1], Mehmet Atıf Durmuş[2], Ahmet Osman Ölçer[2], Hilal Korkut[2], Cem Sevik[3,4], İbrahim Sarpkaya[2], Talip Serkan Kasırga[2+]

[1] Department of Physics, Bilkent University, Ankara 06800 Turkey

[2] Bilkent University UNAM – Institute of Materials Science and Nanotechnology, Ankara, 06800 Turkey

[3] Department of Physics and NANOLab Center of Excellence, University of Antwerp, Groenenborgerlaan 171, B-2020 Antwerp, Belgium

[4] Department of Mechanical Engineering, Faculty of Engineering, Eskişehir Technical University, Eskişehir 26555 Turkey

[*] Equal contribution

[+] Corresponding email: kasirga@unam.bilkent.edu.tr

**Abstract**

Excitons in monolayer transition metal dichalcogenides (TMDCs) offer intriguing new possibilities for optoelectronics with no analogues in bulk semiconductors. Yet, intrinsic defects in TMDCs limit the radiative exciton recombination pathways. As a result, the photoluminescence (PL) quantum yield (QY) is limited. Methods like superacid treatment, electrical doping, and plasmonic engineering can inhibit nonradiative decay channels and enhance PL. Here, we show a more straightforward approach that allows PL enhancement. An engineered vertical van der Waals (vdW) metal-monolayer semiconductor junction (MSJ) results in PL enhancement of more than an order of magnitude at technologically relevant excitation powers. Such MSJ can be constructed by vertically stacking metals with suitable work function either above or below a monolayer semiconducting TMDC. Our experiments reveal that the underlying PL enhancement mechanism is to be the suppressed exciton quenching due to the absence of metal-induced gap states and weak Fermi level pinning, thanks to the vdW gapped interface between the metal and the TMDC. Our time-resolved PL measurements further indicate that reduced exciton-exciton annihilation, even at high generation rates, contributes to the observed PL enhancement. The PL intensity is further increased by the proximity of surface plasmons in the metal with the TMDC layer. Our findings shed light on the interaction at vdW metal-semiconductor interfaces and offer a path to improving the optoelectronic performance of semiconducting TMDCs.

**Main text**

**Introduction**

Strong geometric confinement, a weak dielectric screening, and the direct band gap in two-dimensional (2D) semiconductors of transition metal dichalcogenides (TMDCs) create a unique platform to study exciton physics via photoluminescence (PL) spectroscopy and can impact potential optoelectronics applications. Large binding energies reaching up to 500 meV, the presence of charged excitons, and the ability to form interlayer excitons even at room temperature make monolayer TMDCs better alternatives for excitonic device studies[1–3]. However, the excitonic states enabled by the strong Coulomb interactions result in many nonradiative exciton recombination pathways, such as the conversion of excitonic states to spin- and valley-forbidden dark states. As a result, the PL intensity in monolayer TMDCs is greatly reduced. PL enhancement via superacid treatment to passivate crystal defects[4,5], electrical doping[6], or plasmonic engineering[7] has been reported in the literature. However, these methods involve harsh chemical treatments, fabrication of complex devices, or interfacing materials with plasmonic structures to enhance the PL emission. Also, these methods only work at very low excitation powers, as at high excitations, exciton-exciton interactions lead to non-radiative Auger recombination[8,9]. A simple method is desired to enhance the PL for better optical performance of TMDCs.

At the metal-semiconductor junction (MSJ), the Schottky barrier height (SBH) is a key factor that determines the type and number of injected charge carriers[10]. Many studies have been devoted to understanding the SBH at two-dimensional (2D) transition metal dichalcogenides (TMDC) - metal interfaces as it is both fundamental and applied interest[11–15]. Besides the band alignment between the metal and semiconductor, there are other factors that affect the quality of the MSJ. For instance, when creating an MSJ with a three-dimensional (3D) metal and a 2D semiconductor, direct evaporation of the metal results in defect formation, metal-induced gap states (MIGS)[16], and Fermi-level pinning (FLP)[17] in the semiconductor. Consequently, a single type of charge transfer occurs across the MSJ, and the semiconducting properties of TMDC are heavily hampered at the junction. **Figure 1a** depicts the schematic of a 3D metal-2D semiconductor junction fabricated via conventional metal deposition and its impact on the band configuration at the MSJ. Recent studies have shown how such nuisances can be avoided via controlled evaporation or transfer of metals over to 2D semiconductors[11,12,14,18,19]. **Figure 1b** depicts the formation of a van der Waals MSJ.

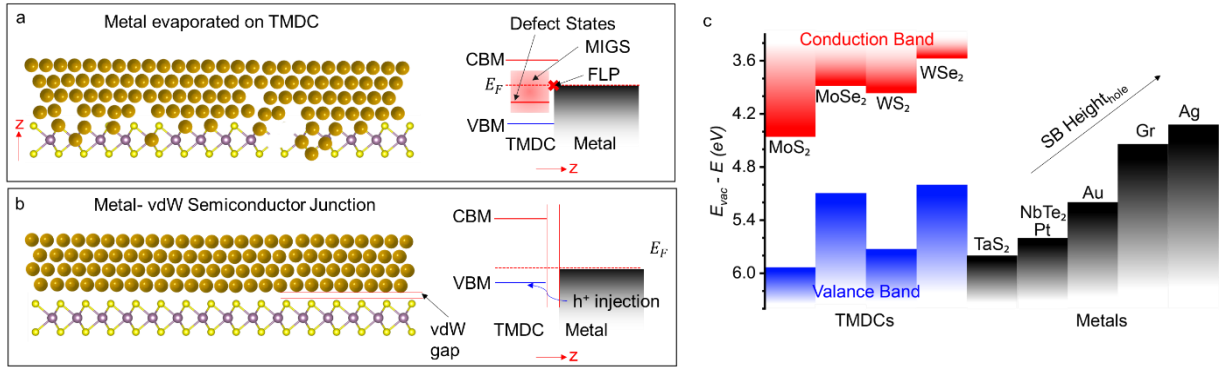

**Figure 1 | Metal-TMDC interface and the band edge alignments. a.** Schematic showing a metal-TMDC interface fabricated via metal evaporation and the consequent band structure at the interface. Metal-induced gap states (MIGS), Fermi level pinning (FLP), and defect states are schematically shown. CBM and VBM stand for conduction band minimum and valance band maximum, respectively. **b.** Schematic showing a metal-TMDC van der Waals (vdW) junction created by transferring the metal. MIGS, FLP, and metal-induced defect states no longer exist. The transfer procedure also leads to a vdW gap. **c.** Schematic showing the work function and band gap alignment of metals and TMDCs used in this study.

Despite the focus of MSJ engineering being on improving the electronic performance of transistor structures[11,12], we propose that a similar approach can be employed to manipulate the excitonic properties of monolayer TMDCs. Although there have been earlier attempts to control the PL energy and intensity of semiconducting TMDCs via metal interfacing, the extent of these studies is different from the present work. In an earlier attempt, Grzeszczyk et al.[20] demonstrated that metallic substrates with properly selected work functions can be used to tune exciton populations in monolayer $MoSe_2$. Recently, it has been shown that small gold disks placed under $WS_2$ monolayers can be used to enhance the PL intensity via hot carrier injection[7]. Similarly, suppressed exciton-exciton annihilation (EEA) due to the screening of dipole-dipole interaction in $WS_2$ monolayers separated from Au by h-BN thin flakes has been shown recently[9].

In this work, we demonstrate that the PL intensity of semiconducting monolayer TMDCs such as $MoS_2$ and $WS_2$ can be enhanced by more than a factor of ten by directly interfacing the material with a metal that allows shallow Schottky barrier for holes, such as $2H-TaS_2$, $2H-NbTe_2$, Au and Pt. We have also demonstrated that PL quenching can also be achieved at will by choosing a metal (Ag, graphene) with a shallow barrier for electrons. We experimentally determine that the reason for the demonstrated enhancement is three-fold: (*i*) The clean van der Waals interface between metal and TMDC results in limited MIGS and weak Fermi level pinning [21,22], which limits non-radiative recombination pathways for excitons, (*ii*) the van der Waals gap limits the free charge transfer from TMDC to the metal or vice-versa and allows large plasmonic field enhancement [12,20,23], and (*iii*) reduced exciton-exciton annihilation due to the screening of the dipole-dipole interactions in the proximity of the metal. To elucidate the underlying physical mechanisms, we performed gate-dependent and time-resolved PL studies. Moreover, we revealed the importance of the van der Waals gap between the metal and the semiconductor to show how metal proximity effects the PL. We supported our findings with temperature-dependent PL experiments and *ab initio* calculations.

**Results and Discussion**

**PL enhancement of metal-semiconductor junctions.** To demonstrate the PL enhancement (and quenching), we used mechanically exfoliated monolayer TMDCs, $MoS_2$, and $WS_2$. We selected 2H-$TaS_2$, 2H-$NbTe_2$, few-layer graphene (Gr), Pt, Au, and Ag as metallic interfaces. Metals are selected based on their work function alignment with the TMDC valance band maxima, as shown in **Figure 1c**. Multi-layers of vdW metals, such as $TaS_2$, $NbTe_2$, and Gr, are obtained via mechanical exfoliation, and ultra-flat 3D metals are thermally evaporated on insulating substrates. Polydimethylsiloxane (PDMS) stamping is used to place TMDC monolayers onto vdW and 3D metals[24] (see Methods for details). To have reference PL measurements, PL and Raman spectra of semiconducting monolayers are collected when they are identified over the PDMS stamp. Also, TMDC crystals are positioned so that they overlap partially with vdW metals to have a reference point for comparing the MSJ and bare TMDC PL intensities. Excitation laser power is closely monitored and kept the same throughout the measurements. PL quantum yield (QY) measurements are calibrated against rhodamine 6G (R6G) embedded in PMMA as the reference following the relevant literature [25]. See Methods for further details of the experimental parameters and procedures.

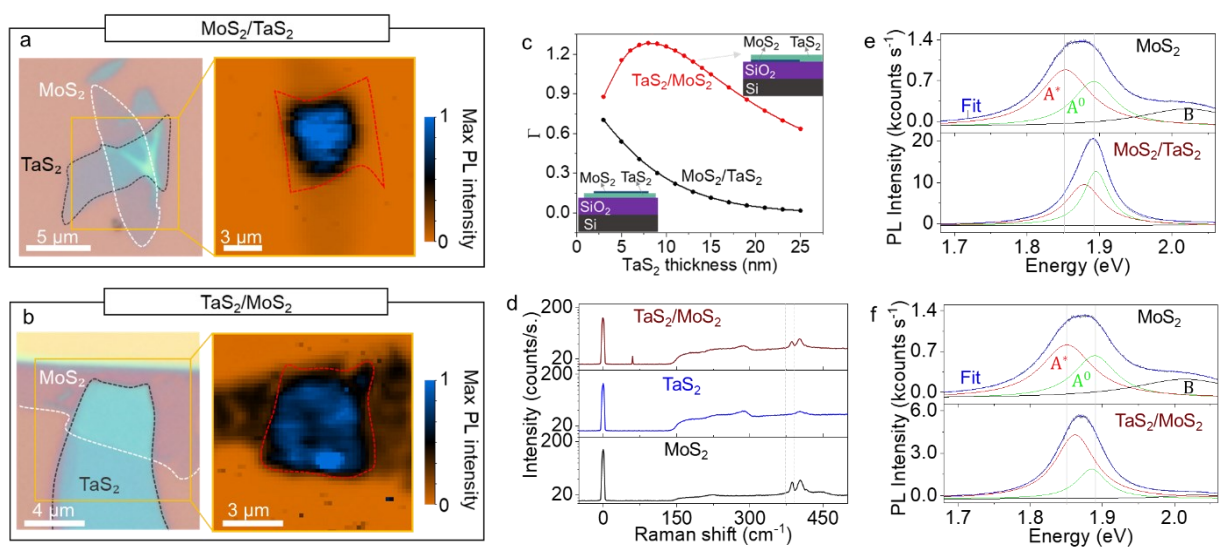

**Figure 2 | PL enhancement with $TaS_2$ above or below $MoS_2$.** Optical microscope images and normalized maximum PL intensity maps of **a.** monolayer $MoS_2$ on 2H-$TaS_2$, **b.** 2H-$TaS_2$ on monolayer $MoS_2$ are shown. The red dashed region in the PL maps marks the heterojunction area. **c.** Optical enhancement factor $\Gamma$ vs. $TaS_2$ thickness when $TaS_2$ is on top (red) and on bottom (black) of the MSJ. **d.** Raman spectra from the MSJ, bare $TaS_2$, and bare $MoS_2$ part of the MSJ. **e.** Pointwise spectra from monolayer $MoS_2$ and $MoS_2/TaS_2$ and **f.** $TaS_2/MoS_2$ are shown. The red, green, black, and blue line fits show the Lorentzian fits for $A^*$, $A^0$, $B$ exciton and the total fit, respectively. All spectra are collected at 150 µW (19 kWcm$^{-2}$).

A couple of optical microscope images and PL intensity maps of $MoS_2$-$TaS_2$ MSJs with PL enhancement are shown in **Figure 2a-b**. PL intensity maps are collected with a near-diffraction limited Gaussian laser spot (spot waist is ~500 nm). Enhancement and quenching in other MSJs are provided in Supporting Information, **Figure S1**. First, monolayer $MoS_2$ on multilayer 2H-$TaS_2$ is investigated (**Figure 2a**). The PL intensity map shows a uniform, 11-fold enhancement in the peak intensity (6.3-fold in integrated PL intensity) over the heterojunction as compared to the bare parts of the $MoS_2$. To show that the enhancement does not originate from thin film interference effects[26], but rather from the interaction between the semiconductor and metal layer, we fabricated $TaS_2$ over $MoS_2$ heterojunctions as shown in **Figure 2b**. The PL intensity map also shows a 4.3-fold enhancement (2.3-fold in integrated PL intensity), indicating that the major contribution is not due to optical interference. To further understand how optical interference through the thin films affects the measured PL intensity, we calculated the thin film optical enhancement factor ($\Gamma$) for the MSJs on the substrate (see Supporting Information for detailed calculations). **Figure 2c** shows $\Gamma$ vs. $TaS_2$ thickness for an MSJ stack on 300 nm thick $SiO_2$ on Si. For $MoS_2/TaS_2$ stack $\Gamma$ is less than one for all $TaS_2$ thicknesses, which means there is suppression of the total emitted PL spectrum. In particular, for the 15-nm $TaS_2$ layer as in the sample in **Figure 2a**, the intrinsic PL enhancement at 1.9 eV should be 90-fold to achieve the measured 11-fold PL enhancement. For TaS$_2$/MoS$_2$ stack, Γ is close to one, which means there is either very small suppression or enhancement due to the optical interference. The suppression of the PL intensity is also confirmed by Raman measurements (**Figure 2d**). Rayleigh and Raman scattering peaks for TaS$_2$/MoS$_2$ stack are suppressed by ~1.45 times which agrees well with calculated $Γ^{-1} = 1.43$. As a result, considering the optical suppression of the emitted signal, the enhancement can be up to two orders of magnitude larger on a suitable MSJ stack.

Another important aspect of the PL emitted from the MSJ is the contribution of various excitonic species to the PL spectrum. **Figure 2e-f** shows pointwise spectra from the bare crystals and the MSJ regions of the samples shown in **Figure 2a-c**. Each PL spectrum is deconvoluted by fitting Lorentzian functions corresponding to neutral exciton ($A^0$), charged excitons ($A^* = A^-$ or $A^+$) and B exciton. Position, full width at half maximum (FWHM), and the integrated area for the peaks are extracted for comparison of different PL spectra. To compare, we defined normalized PL spectral weight for $A^*$ and $A^0$ as:

$$\gamma^i = \frac{I_{A^i}}{I_{Total}}, i = *, 0 \quad \text{(Eq. 1)}$$

where, $I_{A^i}$ and $I_{Total}$ denotes the integrated exciton (neutral or charged) PL intensity and the total PL intensity, respectively. Lorentzian fits to the PL spectra show changes in all excitonic species across bare MoS$_2$ and MSJ. There is a slight blueshift in $A^0$ energy, 4.5±0.5 meV, and a larger blueshift in $A^*$ and B exciton energies by 11.0±0.5 meV and 17.7±1.1 meV, respectively, in TaS$_2$/MoS$_2$ stack. The PL spectral weights, $\gamma_{A^0}$ increases from 0.31 in the monolayer to 0.51 in the MSJ while $\gamma_{A^*}$ increases from 0.44 in the monolayer to 0.48 in MSJ. On the other hand, for the MoS$_2$/TaS$_2$ stack, $A^0$ and $A^*$ energies blueshift by 3.3±0.4 meV and 26.1±0.9 meV, respectively. The relatively large shift in the trion energy might be related to the screening of charged interfacial defects in SiO$_2$, provided by the TaS$_2$ layer. The PL spectral weights, $\gamma_{A^0}$ decreases from 0.28 in the monolayer to 0.24 in the heterostructure while $\gamma_{A^*}$ increases from 0.42 in the monolayer to 0.70 in MSJ. The spectral weight of B exciton decreases from 0.30 to 0.06 in the MSJ. These results indicate that depending on the configuration, the generation rate of neutral and charged excitons increases, overall, in MSJs as compared to the bare MoS$_2$, and there is a blueshift in all spectra. However, as discussed later in the gate dependence studies, we consider that the SiO$_2$ substrate plays a role in the observed energy shifts as HfO$_2$ passivated SiO$_2$ substrates show redshift for $A^0$ and blueshift for $A^*$. Finally, the FWHM values for both $A^0$ and $A^*$ peaks become sharper by ~20 meV in both cases.

These observations show that the observed PL enhancement originates from the interaction between MoS$_2$ and the metallic 2H-TaS$_2$ layer and not by optical interference effects. There are a few mechanisms that can lead to an overall increase in the generation rate of both neutral and charged excitons without significantly altering the binding energies of the excitons. Hole transfer from the metallic layer to the semiconductor could be a candidate for such PL enhancement mechanisms, even at high laser intensities. A similar PL enhancement in semiconducting quantum dots based on the charge transfer from metal nanoparticles has been recently demonstrated[27]. Next, we investigate electrically connected samples to understand the underlying physics of the observed PL enhancement.

**Un-pinned Fermi Level and Field-Effect Dependent PL of Metal-Semiconductor Junction.** First, we fabricated a two-terminal device to assess the charge transfer across the MSJ formed by monolayer MoS$_2$ and 2H-TaS$_2$. We measured the MSJ via a two-terminal device with indium needle contacts[28] (**Figure 3a**). For the current-voltage (IV) measurements the MoS$_2$ side is grounded. The MSJ exhibits a diode-like behaviour with an ideality factor ($n$) of 4.6±0.2 estimated for a small forward bias (0.01 – 0.35 V) and a rectifying ratio of ~50 at ± 1 V. The ideality factor is estimated by fitting to the Schottky diode equation:

$$I = I_L(e^{\frac{qV}{nk_BT}} - 1) \quad \text{(Eq. 2)}$$

where, $I$, $V$, $k_B$, $T$, $q$, and $I_L$ denote drain current, voltage, Boltzmann constant, measurement temperature, unit charge and reverse leakage current, respectively. The ideality factor and the rectifying ratio are comparable to the junctions formed by healed MoS$_2$/1T'-MoTe$_2$ [29] and transferred metal junctions without gating[11]. Once the MSJ is illuminated by a diffraction-limited laser beam at 11.2 µW (1.4 kW cm$^{-2}$), entire current response increases by a factor of 5. IV characteristics of the MSJ indicate a vdW junction with qualities close to those of the engineered junctions. As a result, we conclude that the clean interface between the two vdW materials led to a limited number of defects and reduced metal-induced mid-gap states in the MoS$_2$ monolayer. Thus, the Fermi level of the MoS$_2$ is not pinned to a value, and the Schottky barrier height for the holes or electrons can be freely tuned. This may lead to the charge transfer from/to the metal upon generation of excitons.

Electric field-dependent PL modulation can provide insight into the enhanced PL observed at the MSJ. We fabricated a TaS$_2$/MoS$_2$ partial heterojunction on a 50 nm HfO$_2$ passivated SiO$_2$ (300 nm)/Si substrate. The electric field is applied through the Si back gate, and the heterostructure is grounded through the TaS$_2$ crystal via an indium needle contact[28] (**Figure 3d-e**). **Figure 3f** shows the PL spectra from the MSJ and bare MoS$_2$, collected at 105 µW (14 kW cm$^{-2}$) excitation power at different gate voltages from -100 to 100 V. When an electric field is applied, we observe fundamentally different PL responses from bare and MSJ regions of the sample (**Figure 3f-g**). While the bare monolayer MoS$_2$ follows what is reported in the previous studies[1], PL spectra from the MSJ show only slight intensity and peak energy modulation. **Figure 3g** shows the difference between exciton and trion energies as a function of the applied gate voltage. The inset in **Figure 3g** shows the Fermi energy vs the exciton-trion energy difference and the intercept determines the trion binding energies, 20±2 meV and 19±2 meV binding energy for trions in bare MoS$_2$ and MSJ, respectively[1]. The Fermi energy is calculated by following Ref.[1]. Moreover, $\gamma^*$ and $\gamma^0$ can be strongly modulated in bare MoS$_2$, as shown in **Figure 3h**, while no change is observed in MSJ up to 200 V (see Supporting Information). This is a clear indication of the tight binding of excitons in the MSJ and consistent with the previous reports[1,2].

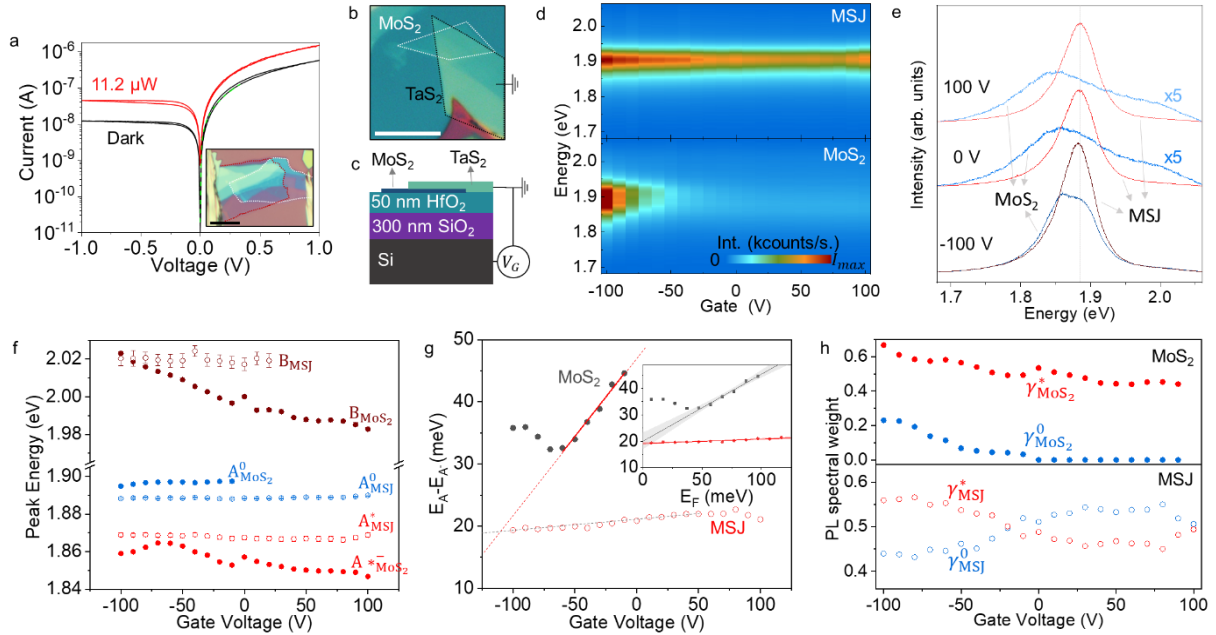

**Figure 3 | Gate-dependent PL of MoS$_2$ and MSJ. a.** IV curve of TaS$_2$/MoS$_2$ metal-semiconductor junction under dark and when the junction is illuminated by a 11.2 µW at 532 nm diffraction limited laser spot parked on the junction. The green dashed curve shows the fit to the Schottky diode equation. Inset shows the optical microscope image of the device. The red dashed line marks the MoS$_2$ and the white dashed line marks the TaS$_2$. The scale bar is 10 µm. The monolayer MoS$_2$ is the bottom part of the entire crystal. **b.** An optical microscope image of a sample device used for gating measurements. An MoS$_2$ monolayer is placed on a HfO$_2$ coated SiO$_2$/Si substrate and partially covered by a TaS$_2$ flake. TaS$_2$ is grounded for gating measurements. The scale bar is 10 µm. **c.** Cross-sectional schematic of the gating junction. The bare part and the MSJ of the sample are illustrated. **d.** Gate vs. photoluminescence intensity map for MSJ and bare MoS$_2$ is shown. The intensity map is shown as kilo-counts/sec. $I_{Max}$ is 3 kcounts/sec. for MSJ and 1.8 kcounts/sec. for bare MoS$_2$. **e.** Spectra at 100, 0 and -100 V are shown for MoS$_2$ and MSJ. The distinct difference between the gate mouldability of the spectra is eminent. **f.** Gate dependence of the energy of $A^*$, $A^0$ and $B$ excitons are shown. While there is little change in MSJ, peak energy of MoS$_2$ is strongly modulated. **g.** The peak energy difference between trion and exciton is plotted against the gate voltage. The Fermi energy ($E_F$) plot in the inset shows 20±2 meV and 19±2 meV binding energy for trions in bare MoS$_2$ and MSJ, respectively. **h.** PL spectral weight vs. the gate voltage is shown for bare MoS$_2$ and the MSJ. The spectral weight of trions and excitons are modulated significantly in the case of MSJ via gate voltage.

The difference between the electric field modulation of MSJ and bare $MoS_2$ can be explained by the ground plane provided by the $TaS_2$ layer. Due to the presence of the ground plane, when an exciton (or trion) is formed, an image dipole forms in the metallic ground layer. As a result, a double dipole configuration is formed with the minimum energy configuration when the dipoles point in opposite directions, parallel to each other[30–33]. The effect of the external displacement field is to rotate the real exciton, however, this requires an increase in the energy of the configuration, as opposed to the bare case since the image exciton is also rotated. As a result, the peak energy of the excitons is much less sensitive to the external displacement field. The same line of reasoning can also be used to understand the spectral weight change of neutral and charge excitons due to the gate voltage. The dipole moment of trions and excitons are different[34], and this difference accounts for the change of the spectral weight as a function of gate voltage. However, a more rigorous calculation of this process is required for an in-depth understanding.

**Effect of van der Waals Gap on PL Enhancement and Time-Resolved PL Measurements.** It is quite peculiar that the free charge carriers in the metallic layer do not quench the PL by introducing excess free charges to the semiconducting layer[35]. We consider that the van der Waals (vdW) gap plays a role. To elucidate that role in the PL enhancement, we compared the PL intensity, the apparent thickness of the monolayers, and the Raman peak intensities of $WS_2$ before transferring onto an Au substrate, before and after thermal annealing the $WS_2$/Au structures. The annealing is performed under an inert ambient, with a slow heating and cooling rate to avoid thermal shock-induced stress on the crystals (details are given in the Methods section). The idea is to remove interfacial adsorbates between the gold and $WS_2$ to reduce the vdW gap height and eliminate the PL enhancement mechanisms. **Figure 4a** shows the optical microscope image of a $WS_2$ monolayer on Au, transferred via PDMS stamping. **Figure 4b** shows the PL intensity map, where intensity counts reach up to 11 kcounts/s. AFM height trace map (**Figure 4c**) shows that the apparent height of the monolayer is ~1.25 nm, which is thicker than monolayers reported in the literature. However, the step from the monolayer to the trilayer part of the crystal is in line with the expected height ~1.4 nm (**Figure S10**).

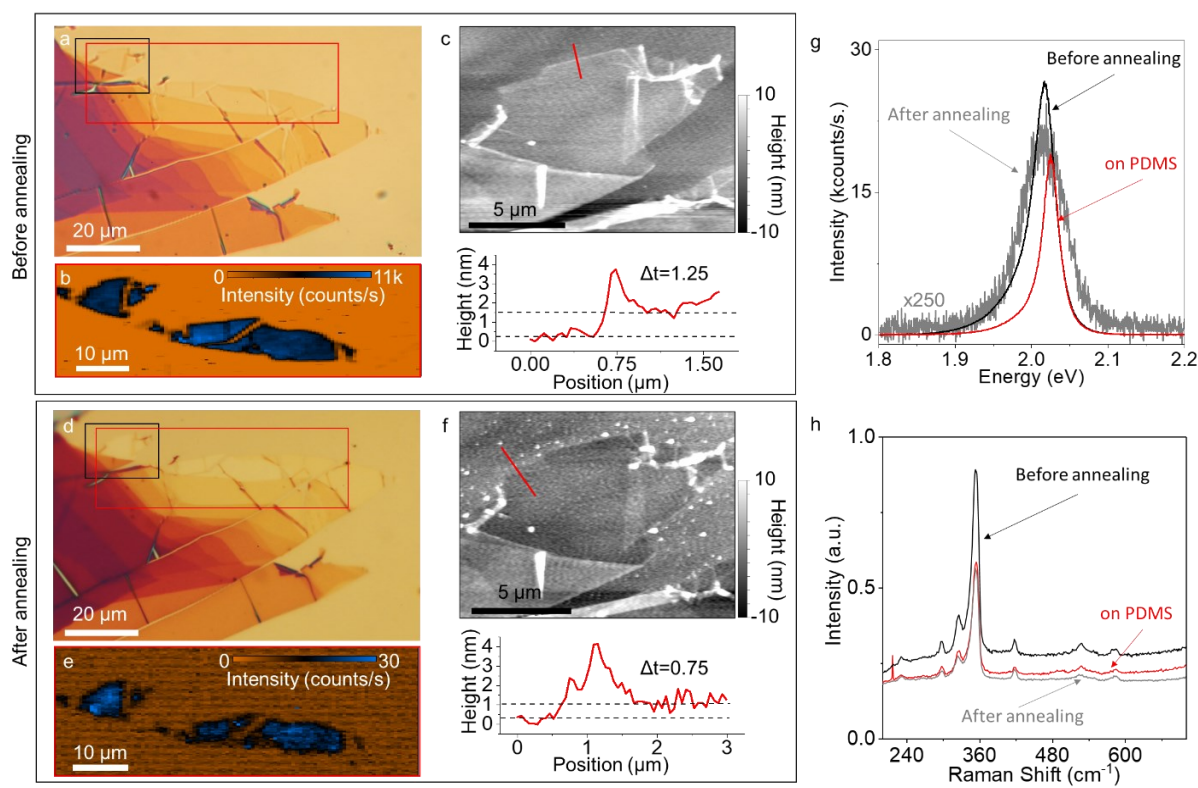

**Figure 4 | Effect of thermal annealing on the van der Waals gap. a.** Optical microscope image of a $WS_2$ crystal on a gold surface after transfer, before thermal annealing, is shown. Red and black rectangles show the regions of PL and AFM maps, respectively. **b.** PL intensity map shows the spatial distribution of the PL maxima on the sample before annealing. **c.** AFM height trace map shows the monolayer and the lower panel shows the height trace from the red line on the map. The thickness of the flake is thicker than what is expected of a monolayer. **d.** Optical microscope image after annealing

is shown. The optical contrast of the monolayer regions is different than what is seen in **a**. Red and black rectangles show the regions of PL and AFM maps, respectively. **e.** PL intensity map shows a dramatic reduction of the counts/s throughout the monolayer. **f.** The height trace map of the same area now shows a 0.75 nm thickness, which is expected for a monolayer $WS_2$. **g.** PL point spectra are shown for the crystal before transferring while on PDMS, after transferring on the Au surface, and after annealing on the Au surface. There is more than 250 times reduction in the intensity after annealing. **h.** Raman spectra for the crystal before transferring, while on PDMS, after transferring on Au surface, and after annealing on Au surface are given in the graph. There is a 1.5 times increase in the PL intensity after transferring $WS_2$ on Au surface. However, after annealing the Raman peak intensities reduce to before transfer values. Also, none of the Raman peaks are affected by the annealing process. All PL measurements are collected with 532 nm laser at 1.7 µW (220 W cm$^{-1}$). 470 µW is used for Raman measurements.

After annealing the sample, the optical contrast of the monolayer regions changes dramatically (**Figure 4d**). Similarly, the PL intensity maximum drops to 30 counts/s, which is almost a 300 times reduction in the mapped area. Moreover, the apparent AFM height trace shows ~0.75 nm thickness for the same monolayer area. These changes show that the vdW gap between the Au and the $WS_2$ flake is smaller after annealing. **Figure 4g** shows a comparison between pointwise PL spectra when the $WS_2$ monolayer is on PDMS, after stamping on Au and after the annealing. There is a 1.5 times increase after the transfer of the flake, yet almost 300 times decrease after annealing. The presence of the van der Waals gap before annealing is also consistent with the recent theoretical work that shows the Au surface can stay about 0.2 nm away from the TMDC surface due to its weak interactions with TMDCs[36].

To assess the effect of thermal annealing on the flake, we performed Raman spectroscopy when $WS_2$ monolayer is on PDMS stamp, after stamping on Au and after the annealing (**Figure 4h**). After transfer, the Raman intensity of the peaks increases by 1.6 times. This can be attributed to the enhanced reflection from the Au substrate as compared to the transparent PDMS. Once the sample is annealed, the Raman intensity drops back to the same intensity for the $WS_2$ on PDMS case. There are no other changes in the Raman spectra. This shows that thermal annealing did not cause any structural change or damage to the crystals. As the vdW gap is closed, the excitation reflected from the gold surface doesn't interact with the sample to produce an enhanced Raman signal. We performed X-ray photoelectron spectroscopy (XPS) to assess the atomic binding energies of $WS_2$ on Au before and after thermal annealing (**Figure S11**). After the annealing, the W 4f binding energy shifts by 0.4 eV, which is consistent with the charge transfer from Au to $WS_2$ because of closing of the vdW gap[37]. Moreover, temperature dependent PL measurements show that below 250 K, the vdW gap shrinks and the enhancement disappears reversibly (**Figure S12**). All these findings are in line with the effect of the vdW gap on the observed PL enhancement mechanisms.

To differentiate the metal-induced PL enhancement and the metallic proximity effects, we placed a 40 nm thick h-BN layer between the Au substrate and the monolayer $WS_2$ flake (**Figure 5**). A similar sample structure has been studied by Lee *et al.*[9], recently. They show that at the low laser power regime (~10 nW), exciton-exciton annihilation and thin film interference effect causes the observed PL enhancement. Here, we focus on how h-BN affects the three mechanisms we proposed as the source of the PL enhancement. The PL spectra of monolayer $WS_2$ on PDMS are used as the reference. When $WS_2$ is placed on 40 nm thick h-BN/Au, we calculated that due to the interference effect there is a factor of ~5 enhancement in the PL as compared to the PDMS-supported case (see Supporting Information). Our PL measurement shows an 8.9 times enhancement of the PL signal (**Figure 5b**). Lorentzian fits to the PL spectra show that both PDMS and h-BN/Au supported monolayers exhibit very similar spectral weight of trions: $\gamma^*_{\text{PDMS}} = 0.09$ and $\gamma^*_{\text{hBN/Au}} = 0.11$ . Moreover, $A^0$ peak shifts from 2.022 eV to 2.006 eV, by 16 meV, on h-BN/Au supported $WS_2$ as compared to the PDMS-supported case. Trion peak also redshifts by 23 meV amount from 1.989 eV to 1.966 eV, after transferring on to h-BN/Au from PDMS.  However, when monolayer $WS_2$ is directly interfaced with the Au, the enhancement is reduced to 3-fold, and the trion spectral ratio has changed significantly from $\gamma^*_{\text{PDMS}} = 0.10$ to $\gamma^*_{\text{Au}} = 0.59$. The change in the spectral weight show that there is a significant charge transfer which results in a dramatic increase of the charged excitons. Also, it is noteworthy that there is no shift in the $A^0$ energy (2.020 eV) but a significant shift in $A^*$ energy from 1.990 eV to 2.005 eV. The blueshift in the trion energy indicates that the binding energy of the trion is decreased when directly interfaced with the Au substrate. Similar results are obtained for different samples (i.e., the sample reported in **Figure 4**). The shift in the trion energy could result from a large number of hole injections from the Au substrate, as the Schottky barrier height for the holes is shallow for $WS_2$. As a result, the trions we observe could be positively charged. However, since the sign of trions is not directly relevant to our discussion, we defer experiments pertaining to the trion sign to future studies.

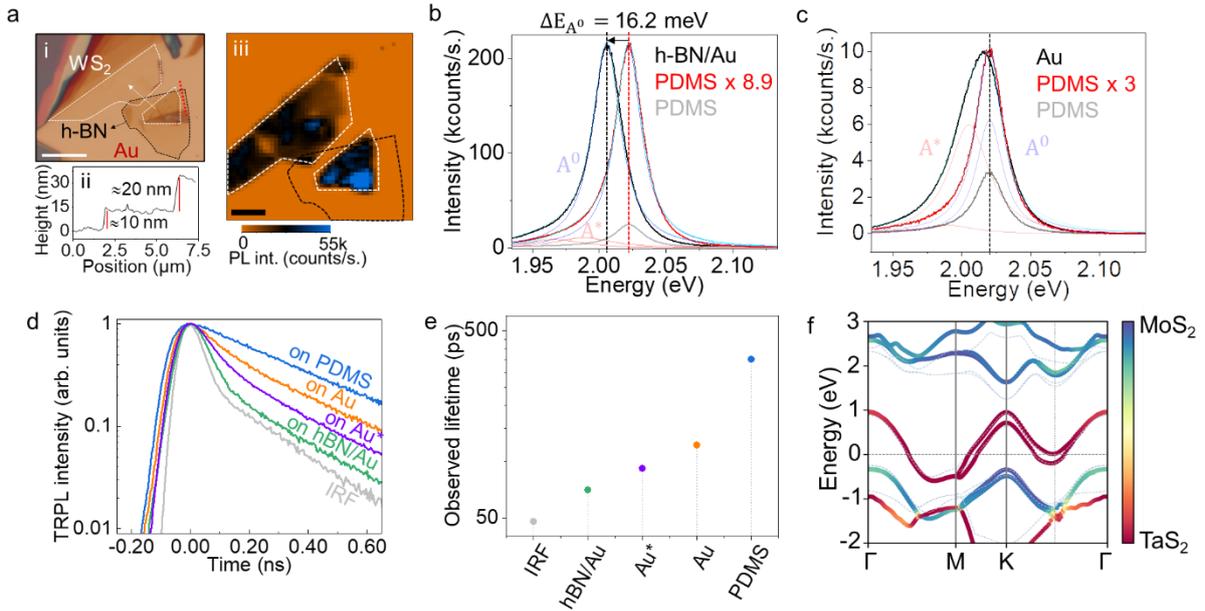

**Figure 5 | Time-resolved PL measurements and the calculated band structure. a. i.** An optical microscope image of a monolayer WS$_2$/h-BN/Au heterostructure is shown. Some parts of the WS$_2$ flake are directly on the Au and some other parts are on the h-BN/Au stack. White and black dashed lines outline the WS$_2$ and h-BN, respectively. **ii.** The red dashed line in **i** shows the AFM height trace along the h-BN layer. The thinner part of the flake is ~10 nm and the thicker part is ~30 nm. **iii.** Maximum PL intensity map and the outlines of the flakes are shown. The part that corresponds to thicker h-BN shows larger enhancement, in line with the optical interference calculations. **b.** PL spectra collected from the WS$_2$ crystal while it is on PDMS and after stamping on h-BN/Au substrate are shown. There is an 8.9-fold enhancement of the overall PL intensity after the transfer and a redshift of the $A^0$, $A^*$ peaks by 16.2±0.2 meV and 22.9±1.5 meV, respectively. The spectral weight of neutral excitons changes slightly upon transfer from $\gamma^0_{PDMS} = 0.91$ to $\gamma^0_{hBN/Au} = 0.89$. **c.** WS$_2$ on gold shows a 3-fold enhancement in the PL intensity after transferring directly over the Au substrate. Moreover, spectral weight neutral excitons change dramatically from $\gamma^0_{PDMS} = 0.91$ to $\gamma^0_{Au} = 0.41$ with no shift in the peak energies. **d.** TRPL intensity vs. time graphs for WS$_2$ on different substrates and the instrument response function are shown. The purple line marked with Au* shows after annealing. **e.** Observed PL lifetime is plotted for different substrates. **f.** *ab initio* band diagram calculations for the MoS$_2$-TaS$_2$ heterostructure with non-relaxed lattice constants. The color bar represents the layer contributions to the corresponding electronic state. The dashed lines and symbols represent the G$_0$W$_0$ and PBE calculations, respectively.

Time-resolved PL (TRPL) measurements can provide insight into the exciton decay channels and illuminate the PL enhancement mechanisms. First, we performed TRPL measurements on monolayer WS$_2$ on PDMS stamp, then after the crystal is transferred to Au and h-BN/Au substrate as well as after annealing the sample to close the vdW gap. The same laser power, 7.5 µW (240 Wcm$^{-2}$), is used for all the measurements. **Figure 5d** shows the normalized TRPL time traces for all three substrates as well as the instrument response function. The extracted PL lifetimes are found to be 352±1, 123±2, and 71±2 for the PDMS, Au and h-BN/Au, respectively (**Figure 5e**). As we performed the TRPL measurements at relatively large laser powers, exciton-exciton annihilation is a significant factor in PL decay. However, after annealing WS$_2$ on Au substrate the observed PL lifetime shortens. This shows that the reduction of the vdW gap between the Au and the WS$_2$ monolayer shortens the PL lifetime as a result of increased charge transfer from the gold substrate.

Finally we would like to discuss that recently, da Jornada *et al.*[23] showed that in atomically thin quasi-two-dimensional metals, broken translational symmetry leads to interband screening via *ab initio* calculations supported by effective model analysis. As a result, they predicted that long-lived plasmons can lead to field intensity enhancement, reaching up to 10$^7$ in monolayer TaS$_2$. Such large field enhancement can be responsible for the observed PL intensity increase in the presence of a van der Waals gap. Despite the fact that this field enhancement might be weaker in the case where 3D metals such as Au are interfaced with the semiconductor, other effects may still lead to the PL enhancement we observed. To further develop an understanding of the band alignment in MSJs, we performed *ab initio* calculations with the GW correction.

The calculated band-gap value of 2.571 eV at the high-symmetry K point for monolayer $MoS_2$ agrees well with prior calculations[38] (**Figure S13**). Due to the substantial mismatch between the equilibrium in-plane lattice constants of $TaS_2$ and $MoS_2$, we initially examine the heterostructure of $MoS_2$-$TaS_2$ structure with $MoS_2$ monolayer's in-plane lattice parameters, while allowing all ions to relax along out-of-plane directions. Surprisingly, the direct GW band gap corresponding to the $MoS_2$ layer reduces to[39] 1.965 eV (**Figure *5f***). Given this result, one might anticipate a significant redshift in the direct exciton energy, yet we experimentally only observe a few meV shifts in photoluminescence (PL) measurements. Generally, we would expect a noticeable alteration in the semiconductor layer due to the metallic screening around the Fermi level induced by the metallic substrate, whereas the observed reduction is significant. This result agrees with what we observe is due to the van der Waals gap and weak interaction between the metal and the semiconducting layers. Another possibility is the invalidity of the plasmon-pole approximation employed in our GW calculations. While this method typically performs well for semiconductors, its application to our MSJs may underestimate the band gap of the $MoS_2$ layer. Energy band diagrams are provided in the supporting information.

**Conclusions**

To summarize, we showed that the vdW gap between metal and semiconductor junctions can be utilized to enhance the PL intensity up to an order of magnitude. The underlying physical mechanisms are determined experimentally. Suppressed exciton quenching due to the absence of metal-induced gap states and weak Fermi level pinning, thanks to the van der Waals gapped interface between the metal and the TMDC, along with reduced exciton-exciton annihilation plays a role in the observed PL enhancement. Moreover, limited charge transfer from metal to semiconductor due to the van der Waals gap results in an alteration of the charged exciton population. The PL intensity is further increased due to the proximity of surface plasmons with the TMDC layer. Our work shows a simple way to enhance the PL via interfacing two-dimensional materials with metals of suitable work functions. Findings here can be used to engineer simple heterojunctions for enhanced optical performance of TMDCs and possibly other ultra-thin luminescent materials.

**Methods**

**Heterostructure fabrication.** TMDC crystals are mechanically exfoliated from ultra-high quality bulk samples using mechanical exfoliation on an elastomeric substrate (PDMS or Gel-Pak®). Then, the elastomeric substrate is attached to a microscope slide with crystals facing out and, via a micromanipulator and microscope, deterministically placed over the target substrate. Target is either an exfoliated metallic TMDC, evaporated metal, or a semiconducting TMDC, depending on the MSJ configuration. For gate-dependent studies, 50 nm thick $HfO_2$ is coated on $SiO_2$/Si substrate to passivate the $SiO_2$ defects and charged impurities, which leads to inconsistent results and degradation in TMDC properties[28].

**PL and Raman measurements.** PL and Raman measurements are collected using the same setup (Witec alpha 300s) equipped with a 532 nm laser and a Czerny-Turner spectrometer with a Peltier-plate cooled CCD camera. Laser power is monitored after the objective, before each measurement using a calibrated Si photodiode by Thorlabs.

**Quantum yield measurements.** For absolute quantum yield (a-QY) measurements, we used rhodamine 6G (R6G) dye dispersed in PMMA thin film following the literature[25]. We performed measurements using an integrating sphere and calibrated the emission from R6G via an integrating sphere. Then, we determined the absorption of TMDC film and the Reference under the Raman/PL microscope. Finally, the PL intensity of the reference R6G film is measured under the same microscope. Quantum yield calibration and results are shown in Supporting Information.

**Thermal annealing of $WS_2$/Au MSJ.** Thermal annealing is performed under atmospheric pressure in an Ar gas environment at 250 °C for 3 hours. The furnace is ramped at 5 K minute$^{-1}$ from room temperature to the annealing temperature and cooled down naturally at a similar rate. The annealed samples are characterized for any optical, structural, and compositional defects. No samples show measurable signs of strain or change in composition after annealing.

**DFT calculations.** The single-particle wave functions and corresponding energies are acquired through DFT using Quantum ESPRESSO (QE)[40], employing Perdew-Burke-Ernzerhof (PBE)[41] norm-conserving, fully relativistic pseudopotentials within the generalized gradient approximation (GGA)[42] from PseudoDojo project[43]. The settings include a plane-wave energy cutoff of 120 Ry a vacuum separation of 60 bohr between periodic repetitions of the simulation cell along the out-of-plane direction, and a symmetrical k-grid of 42×42×1 (corre-sponding to 169 k points, centered at Γ). Additionally, we implemented Grimme's dispersion correction (referred to as Grimme-D2 in QE)[44,45] to account for the van der Waals (vdW) interaction.

The many body perturbation theory analyses, conducted on the basis of DFT generated wavefunctions, were carried out utilizing the YAMBO code[46,47]. The G0W0 corrections[48,49] to the single-particle eigenvalues were determined using the plasmon-pole approximation for dynamic electronic screening. Convergence in both direct and indirect band gaps was achieved by summing over 600 and 600 states for screening and Green's functions, respectively. These corrections were applied to the top 2 valence bands and the bottom 2 conduction bands.

Subsequently, the BSE[50] calculations for excitons was addressed within the Tamm-Dancoff approximation, incorporating RPA static screening summed over 400 bands. The exciton energies and their wave functions were obtained by direct diagonalization of BSE Hamiltonian. The slab technique[38] was used along the out-of-plane direction to eliminate the long-ranged interactions with the repeated periodic images of the systems in both G0W0 and BSE steps.

**Author Contributions**

HMS and AAS contributed equally and performed device fabrication, all PL measurements and analysed the data. AP performed the XPS measurements and analysis, KNK performed the optical enhancement factor calculations. UB performed some gate dependent PL measurements. MAD, AOÖ, KH and İS performed the low temperature and time resolved PL measurements. CS performed *ab initio* calculations. TSK conceived the experiments, performed the data analysis, and wrote the manuscript. All authors commented on the manuscript and discussed the results.

**Acknowledgments**

TSK acknowledges funding from TÜBİTAK under grant no:121F366.

# Supporting Information: Photoluminescence enhancement at the vertical van der Waals semiconductor-metal heterostructures

Hafiz Muhammad Shakir[1*], Abdulsalam Aji Suleiman[2*], Kübra Nur Kalkan[2], Amir Parsi[2], Uğur Başçı[1], Mehmet Atıf Durmuş[2], Ahmet Osman Ölçer[2], Hilal Korkut[2], Cem Sevik[3,4], İbrahim Sarpkaya[2], Talip Serkan Kasırga[2+]

[1] Department of Physics, Bilkent University, Ankara 06800 Turkey

[2] Bilkent University UNAM – Institute of Materials Science and Nanotechnology, Ankara, 06800 Turkey

[3] Department of Physics and NANOLab Center of Excellence, University of Antwerp, Groenenborgerlaan 171, B-2020 Antwerp, Belgium

[4] Department of Mechanical Engineering, Faculty of Engineering, Eskişehir Techical University, Eskişehir 26555 Turkey

* Equal contribution

+ Corresponding email: kasirga@unam.bilkent.edu.tr

1. **Photoluminescence enhancement and quenching in various metal-semiconductor junctions**

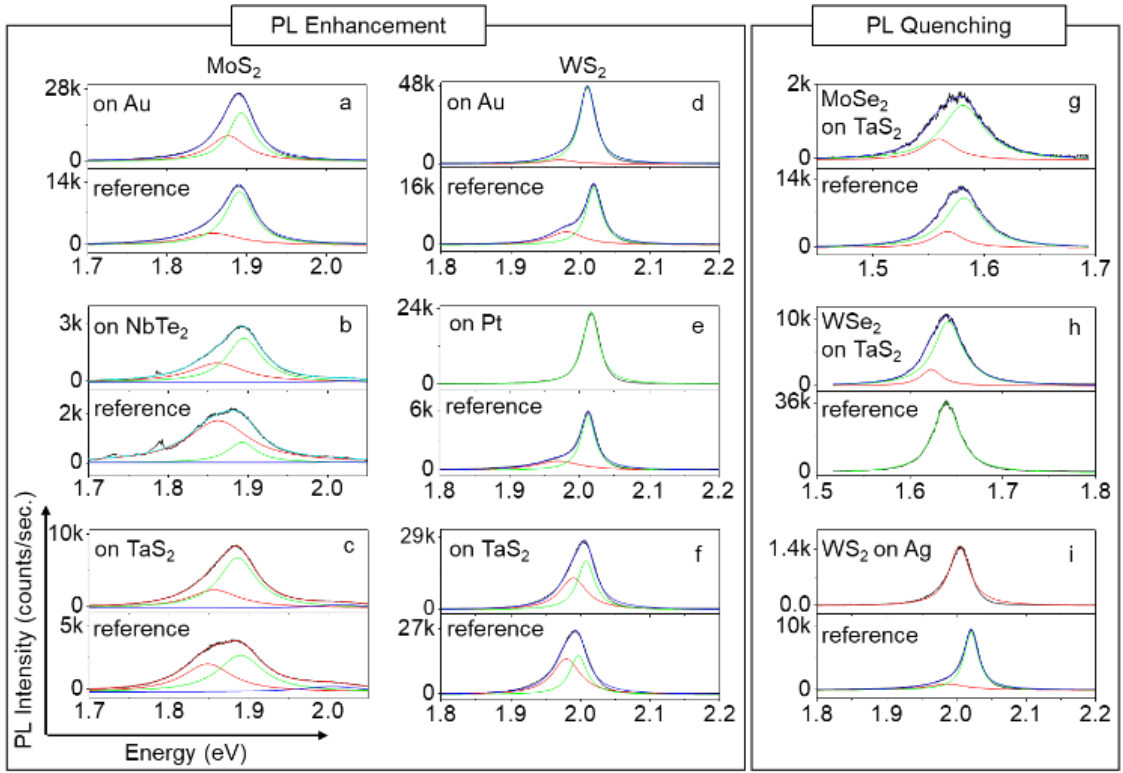

**Figure S1** The left panel shows a collection of PL spectra with enhancement (or no quenching) and the right panel shows a collection of PL spectra with quenching for various TMDC-metal heterojunctions. **a.** $MoS_2$/Au, **b.** $MoS_2$/$NbTe_2$, **c.** $MoS_2$/$TaS_2$, **d.** $WS_2$/Au, **e.** $WS_2$/Pt, **f.** $WS_2$/$TaS_2$, **g.** $MoSe_2$/$TaS_2$, **h.** $WSe_2$/$TaS_2$, **i.** $WS_2$/Ag. For bulk metals, reference PL spectra are collected from the TMDC on PDMS stamp. The red, green, and blue lines are Lorentzian fits corresponding to $A^*$, $A^0$, and $B$, respectively. The sum of the fits is represented by cyan line.

## 2. Optical enhancement factor calculations

In the case for photoluminescence spectrum intensity measurements of heterostructures, optical interference effects must be taken into account to truly understand PL phenomenon[1]. This is because the incident laser beam passing through a material is reflected multiple times inside each layer of the heterostructure. These multiple internal reflections at each interface between layers with different refractive indices affect the PL intensity of the material[1,2].

Our sample is made of $WS_2$ (0.78nm) onto hBN/Au heterostructure with 40nm and 90nm thicknesses, respectively. To estimate PL response of our sample due to optical interference effects, we calculated an enhancement factor by taking the ratio of emission intensities for monolayer $WS_2$ on-substrate and freestanding $WS_2$[3].

### 2.1. Freestanding $WS_2$

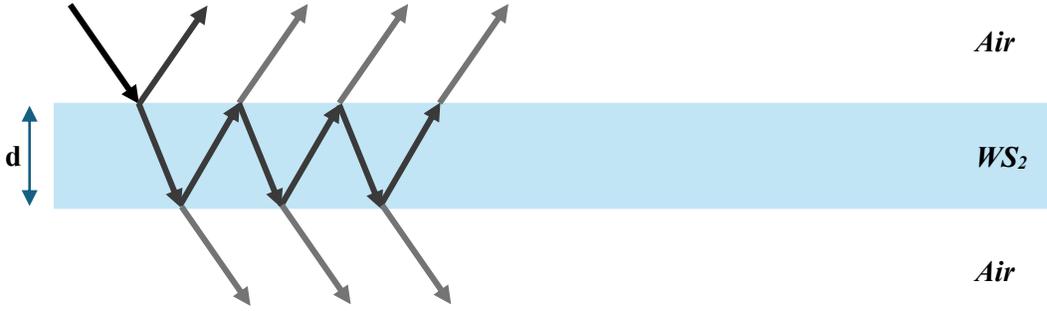

**Figure S2** Freestanding $WS_2$ with thickness d.

### 2.2. Absorption Coefficient Freestanding $WS_2$ (λ = 532 nm):

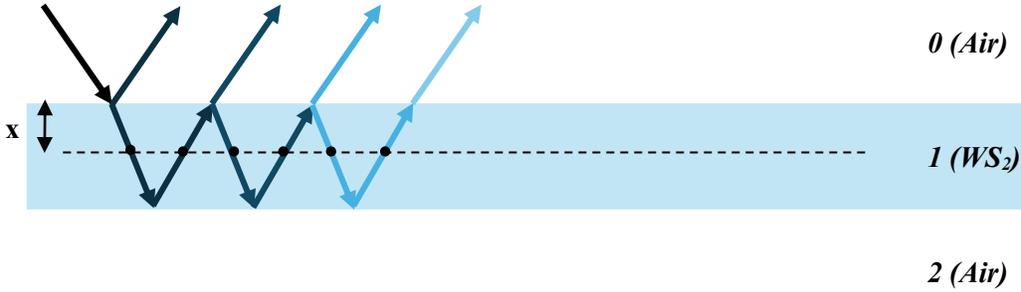

**Figure S3** Absorption of laser beam at a depth x in suspended $WS_2$.

Refractive index of $WS_2$ (medium-1) at λ=532 nm is $n_a(WS_2) \equiv n_1 = 4 - 0.9i$ [2]. As the beam of light passes through any medium, there occurs a geometric phase difference $\beta_i = \frac{2\pi n_i d_i}{\lambda}$

$$\beta_1 = \frac{2\pi n_1 d_1}{\lambda} \quad and \quad \beta_x = \frac{2\pi n_1 x}{\lambda}$$

where $\beta_1$ and $\beta_x$ are geometric phases for medium-1 with thickness $d_1$ and x. Fresnel reflection and transmission coefficients are written to describe reflection and transmission of beam of light travelling from medium-i to medium-j; $r_{ij} = \frac{n_i - n_j}{n_i + n_j}, t_{ij} = \frac{2n_i}{n_i + n_j}$ [3,4]

$$t_{01} = \frac{2n_0}{n_0 + n_1}, \quad r_{01} = \frac{n_0 - n_1}{n_0 + n_1}, \quad r_{12} = \frac{n_1 - n_2}{n_1 + n_2} \, where \, n_0 = n_2 = 1 \, (Air) \qquad 1$$

where $t_{01}$ is the transmission coefficient for the beam from medium-0(Air) to medium-1 (WS$_2$), $r_{01}$ and $r_{12}$ are reflection coefficients for the beam from medium-0(Air) to medium-1 (WS$_2$) and from medium-1 (WS$_2$) to medium-2 (Air), respectively.

To calculate the absorption coefficient at depth x, we consider each line passing through the point at d=x, and we make summation to find the total absorption at depth x [1,5]:

$$a_1 = e^{-i\beta_x}t_{01}$$
$$a_2 = e^{-i\beta_1}t_{01}e^{-i(\beta_1-\beta_x)}r_{12} = t_{01}r_{12}e^{-i(2\beta_1-\beta_x)}$$
$$a_3 = e^{-i\beta_1}t_{01}e^{-i\beta_1}r_{12}e^{-i\beta_x}r_{10} = a_1(r_{12}r_{10}e^{-2i\beta_1})$$
$$a_4 = e^{-i\beta_1}t_{01}e^{-i\beta_1}r_{12}e^{-i\beta_1}r_{10}e^{-i(\beta_1-\beta_x)}r_{12} = a_2(r_{12}r_{10}e^{-2i\beta_1})$$
$$\vdots$$

In general,

$$a_{2n+1} = a_1(r_{12}r_{10}e^{-2i\beta_1})^n, \qquad 2$$

$$a_{2n+2} = a_2(r_{12}r_{10}e^{-2i\beta_1})^n. \qquad 3$$

The total amount of absorption at x can be expressed as a summation of all coefficients defined at d=x

$$F_{abs} = \sum_{n\geq 0}[e^{-i\beta_x}t_{01} + e^{-i(2\beta_1-\beta_x)}t_{01}r_{12}](r_{12}r_{10}e^{-2i\beta_1})^n \qquad 4$$

The second part of the right-hand side in Eqn(4) is written in terms of geometric sum

$$F_{abs} = t_{01}\frac{e^{-i\beta_x} + r_{12}e^{-i(2\beta_1-\beta_x)}}{1 - r_{10}r_{12}e^{-2i\beta_1}}. \qquad 5$$

Recalling Stoke's relations between Fresnel reflection and transmission coefficients [4] ($r_{ij} = -r_{ji}$ and $1 = t_{ij}t_{ji} - r_{ij}r_{ji}$)

$$F_{abs} = t_{01}\frac{e^{-i\beta_x} + r_{12}e^{-i(2\beta_1-\beta_x)}}{1 + r_{01}r_{12}e^{-2i\beta_1}}. \qquad 6$$

### 2.3. Scattering Coefficient Freestanding WS$_2$ ($\lambda$' = 615 nm):

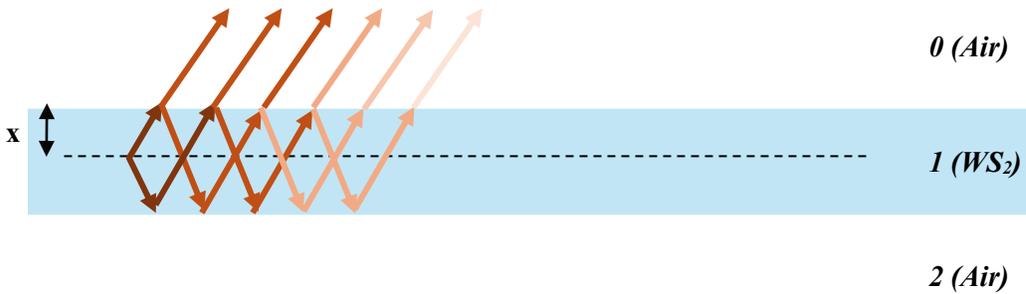

**Figure S4** Scattering of laser beam at a depth x in suspended WS$_2$.

Refractive index of WS$_2$ (medium-1) at λ'=615 nm is $n_s(WS_2) \equiv n_1' = 4 - 1.8i$ [2]. The geometric phase differences at λ'=615 nm for medium-1 with thickness $d_1$ and x

$$\beta_1' = \frac{2\pi n_1' d_1}{\lambda'}, \qquad \beta_x' = \frac{2\pi n_1' x}{\lambda'}.$$

Corresponding Fresnel reflection and transmission coefficients can be written as

$$t_{10} = \frac{2n_1'}{n_1' + n_0}, \quad r_{10}' = \frac{n_1' - n_0}{n_1' + n_0}, \quad r_{12}' = \frac{n_1' - n_2}{n_1' + n_2} \text{ where } n_0 = n_2 = 1 \text{ (Air)}.$$

Similar to the absorption coefficient calculation, we consider each point at depth x and we make summation for the total scattering coefficient at depth x.

$$\begin{aligned}
b_1 &= e^{-i\beta_x'} t_{10} \\
b_2 &= e^{-i\beta_1'} t_{10} e^{-i(\beta_1' - \beta_x')} r_{12}' \\
b_3 &= e^{-i\beta_1'} t_{10} e^{-i\beta_1'} r_{12}' e^{-i\beta_x'} r_{10}' = b_1 \left( r_{12}' r_{10}' e^{-i2\beta_1'} \right) \\
b_4 &= e^{-i\beta_1'} t_{10} e^{-i\beta_1'} r_{12}' e^{-i\beta_1'} r_{10}' e^{-i(\beta_1' - \beta_x')} r_{12}' = b_2 \left( r_{12}' r_{10}' e^{-i2\beta_1'} \right) \\
&\vdots
\end{aligned}$$

$$b_{2n+1} = b_1 \left( r_{12}' r_{10}' e^{-i2\beta_1'} \right)^n \qquad 7$$

$$b_{2n+2} = b_2 \left( r_{12}' r_{10}' e^{-i2\beta_1'} \right)^n \qquad 8$$

By summing over scattering coefficients at depth x, the total amount of scattering coefficient becomes

$$F_{sc} = \sum_{n \geq 0} \left[ t_{10} e^{-i\beta_x'} + t_{10} r_{12}' e^{-i(2\beta_1' - \beta_x')} \right] \left( r_{12}' r_{10}' e^{-i2\beta_1'} \right)^n \qquad 9$$

$$\Rightarrow F_{sc} = t_{10} \frac{e^{-i\beta_x'} + r_{12}' e^{-i(2\beta_1' - \beta_x')}}{1 + r_{10}' r_{12}' e^{-i2\beta_1'}} \qquad 10$$

The last step for freestanding WS$_2$ is to calculate its intensity using both total absorption and scattering coefficients [1]

$$I_{WS_2}^{susp} = \int_0^{d_1} |F_{abs} \cdot F_{sc}|^2 dx \qquad 11$$

### 2.4. On-substrate WS$_2$

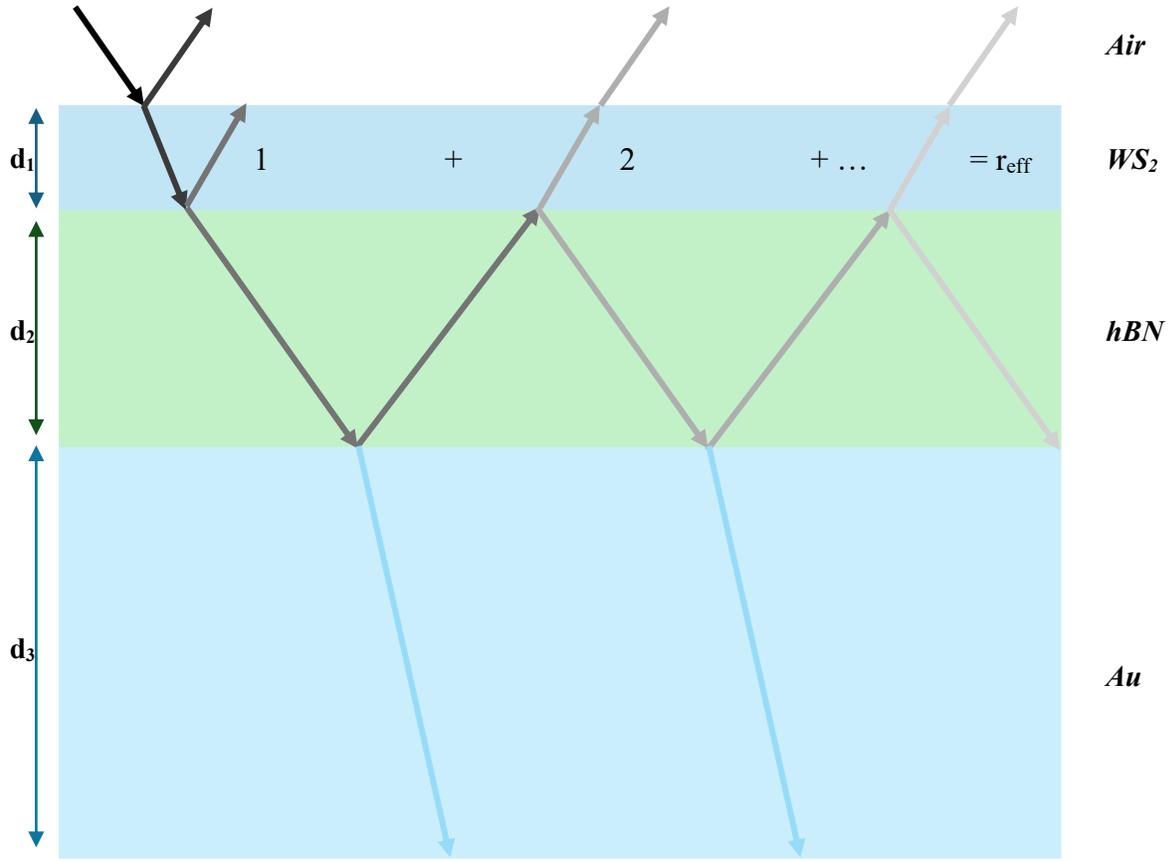

**Figure S5** Schematic diagram of the beam of light passing through $WS_2$ on-substrate hBN/Au.

In the case of $WS_2$ on-substrate (Figure **S5**), considering bottom layers of $WS_2$ as effective medium the effective Fresnel reflection coefficient is derived to define multiple reflections for absorption and scattering coefficients at depth x in medium-1[1]. For this reason, each term is written in terms of corresponding reflection and transmission coefficients including geometric phase differences

$$c_1 = r_{12}$$
$$c_2 = t_{12}e^{-i\beta_2}r_{23}e^{-i\beta_2}t_{21}$$
$$c_3 = t_{12}e^{-i\beta_2}r_{23}e^{-i\beta_2}r_{21}e^{-i\beta_2}r_{23}e^{-i\beta_2}t_{21} = t_{12}t_{21}r_{23}e^{-2i\beta_2}(r_{21}r_{23}e^{-2i\beta_2})$$
$$\vdots$$
$$c_n = c_2(r_{21}r_{23}e^{-2i\beta_2})^{n-2}.$$

Then the effective Fresnel reflection is written in terms of effective coefficients stated above

$$r = r_{12} + t_{12}t_{21}r_{23}e^{-2i\beta_2}\sum_{n\geq 3}(r_{21}r_{23}e^{-2i\beta_2})^{n-2}$$

Recalling geometric sum formulation, Equation 14 becomes

$$r = r_{12} + \frac{t_{12}t_{21}r_{23}e^{-2i\beta_2}}{1 - r_{21}r_{23}e^{-2i\beta_2}}.$$

Similarly, considering reflection and transmission coefficients' relation (Stoke's relations) ($r_{ij} = -r_{ji}$ and $t_{ij}t_{ji} = 1 - r_{ij}^2$) [3,4],

$$r = \frac{r_{12} + r_{23}e^{-2i\beta_2}}{1 + r_{12}r_{23}e^{-2i\beta_2}} \qquad 16$$

where $\beta_2$ is geometric phase difference related to medium-2.

### 2.5. Absorption Coefficient WS$_2$-on-substrate ($\lambda$ = 532 nm):

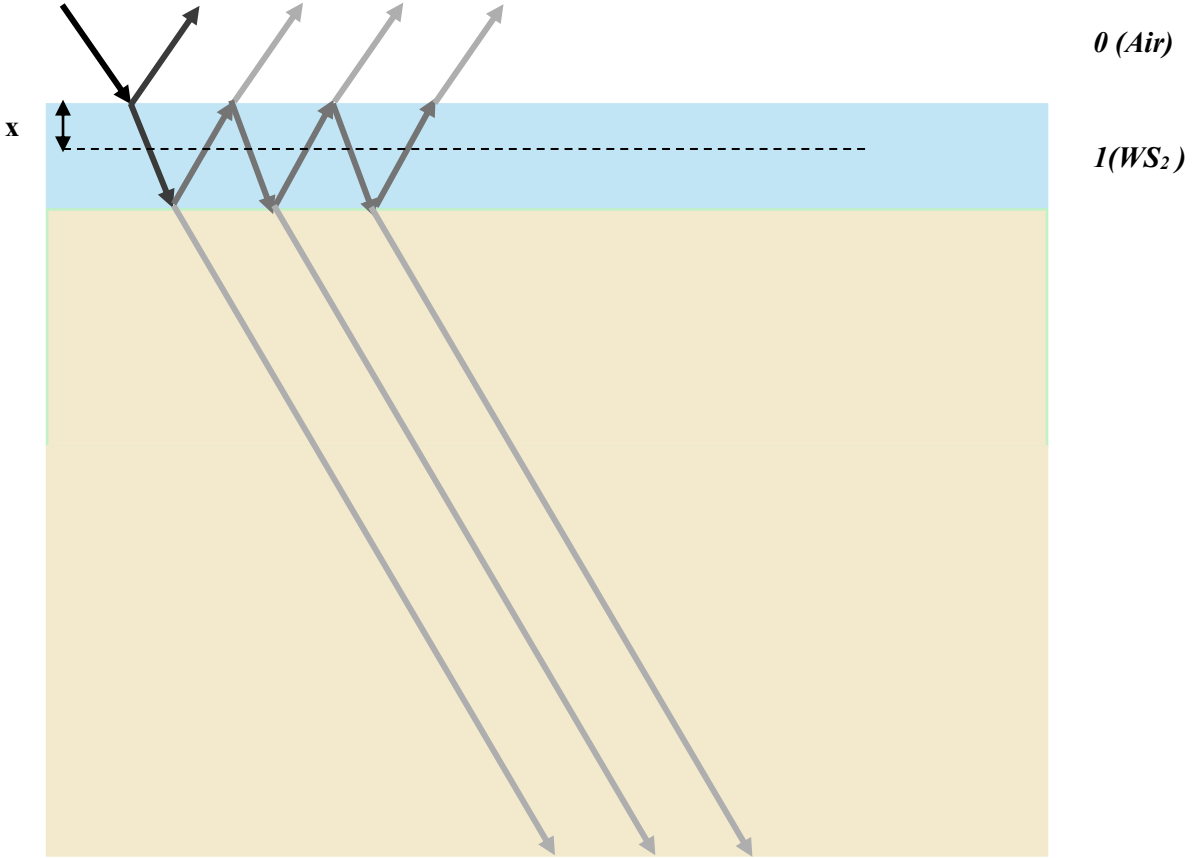

**Figure S6** Schematic diagram of light beam passing through a sample with WS$_2$ (medium-1), hBN (medium-2) and Au (medium-3). Dots on the dashed line represent the points of absorption at depth x.

Refractive indices of WS$_2$, hBN and Au at $\lambda$=532 nm and corresponding geometric phase differences for medium-1 and medium-2 are expressed as

$$n(WS_2)|_{abs} \equiv n_1 = 4 - 0.9i, \qquad n(hBN)|_{abs} \equiv n_2 = 1.85, \qquad n(Au)|_{abs} \equiv n_3 = 0.55 - i2.2,$$

$$\beta_1 = \frac{2\pi n_1 d_1}{\lambda}, \qquad \beta_x = \frac{2\pi n_1 x}{\lambda}, \qquad \beta_2 = \frac{2\pi n_2 d_2}{\lambda}.$$

As we will define the effective Fresnel reflection coefficient for both absorption and scattering cases, we start with the effective reflection for absorption ($\lambda$=532nm) $(r \rightarrow r')$

$$r' = \frac{r_{12}^{abs} + r_{23}^{abs} e^{-2i\beta_2}}{1 + r_{12}^{abs} r_{23}^{abs} e^{-2i\beta_2}} \qquad 17$$

where $r_{12}^{abs} = \frac{n_1 - n_2}{n_1 + n_2}$ and $r_{23}^{abs} = \frac{n_2 - n_3}{n_2 + n_3}$.

Similar to the freestanding case, we take into account each line passing through the point at $d = x$ in order to find total absorption coefficient

$$\begin{aligned}
a_1^{on-subs} &= e^{-i\beta_x} t_{01} \\
a_2^{on-subs} &= e^{-i\beta_1} t_{01} e^{-i(\beta_1 - \beta_x)} r' \\
a_3^{on-subs} &= e^{-i\beta_1} t_{01} e^{-i\beta_1} r' e^{-i\beta_x} r_{10} = a_1^{on-subs}(r' r_{10} e^{-2i\beta_1}) \\
a_4^{on-subs} &= e^{-i\beta_1} t_{01} e^{-i\beta_1} r' e^{-i\beta_1} r_{10} e^{-i(\beta_1 - \beta_x)} = a_2^{on-subs}(r' r_{10} e^{-2i\beta_1}) \\
&\vdots \\
\Rightarrow&
\end{aligned}$$

$$a_{2n+1}^{on-subs} = a_1^{on-subs}(r' r_{10} e^{-2i\beta_1})^n \qquad 18$$

$$a_{2n+2}^{on-subs} = a_2^{on-subs}(r' r_{10} e^{-2i\beta_1})^n \qquad 19$$

It is important to note that $t_{01}$, $r_{10}$, $\beta_1$ and $\beta_x$ are the same as the case for freestanding WS$_2$. For on-substrate WS$_2$, the total amount of absorption

$$F_{abs}^{on-subs} = \sum_{n \geq 0} [t_{01} e^{-i\beta_x} + t_{01} r' e^{-i(2\beta_1 - \beta_x)}](r' r_{10} e^{-2i\beta_1})^n \qquad 20$$

$$F_{abs}^{on-subs} = t_{01} \frac{e^{-i\beta_x} + r' e^{-i(2\beta_1 - \beta_x)}}{1 + r' r_{10} e^{-2i\beta_1}} \qquad 21$$

### 2.6. Scattering Coefficient WS$_2$-on-substrate (λ' = 615 nm):

Refractive indices of WS$_2$, hBN and Au at λ'=615 nm and corresponding geometric phase differences for medium-1 and medium-2 are written as

$n(WS_2)|_{sc} \equiv n_1' = 4 - 1.8i$, $n(hBN)|_{sc} \equiv n_2 = 1.85$, $n(Au)|_{abs} \equiv n_3' = 0.23 - i3.1$, [2]

$$\beta_1' = \frac{2\pi n_1' d_1}{\lambda'}, \qquad \beta_x' = \frac{2\pi n_1' x}{\lambda'}, \qquad \beta_2' = \frac{2\pi n_2 d_2}{\lambda'}.$$

The effective Fresnel reflection for scattering $r''$ $(r \to r'')$[1]

$$r'' = \frac{r_{12}^{sc} + r_{23}^{sc} e^{-2i\beta_2'}}{1 + r_{12}^{sc} r_{23}^{sc} e^{-2i\beta_2'}} \qquad 22$$

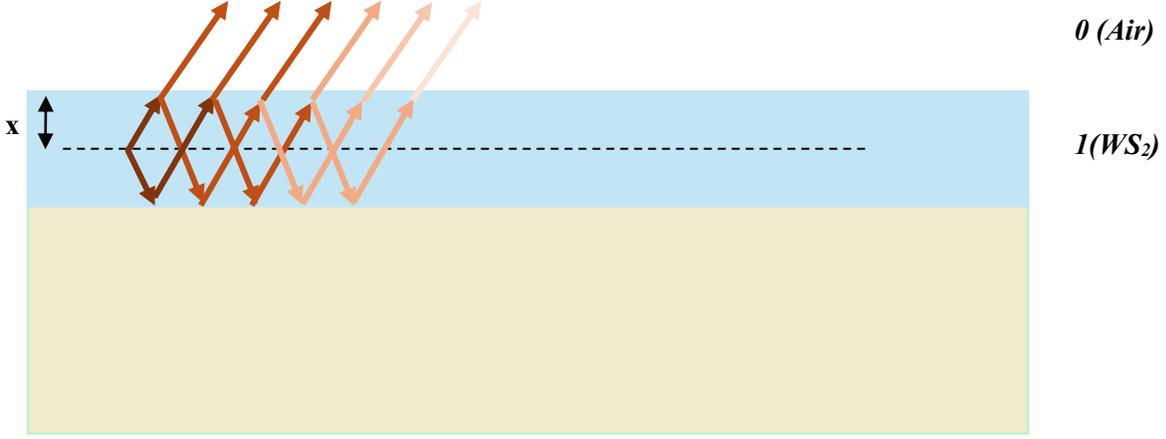

**Figure S7** Scattering of laser beam at a depth x in WS$_2$ on hBN/Au substrate. hBN/Au substrate is considered as the effective medium.

Total amount of scattering at depth x:

$$b_1^{on-subs} = e^{-i\beta'_x}t_{10}$$
$$b_2^{on-subs} = e^{-i(\beta'_1-\beta'_x)}r''e^{-i\beta'_1}t_{10}$$
$$b_3^{on-subs} = e^{-i\beta'_1}t_{10}e^{-i\beta'_1}r''e^{-i\beta'_x}r'_{10} = b_1^{on-subs}(r''r'_{10}e^{-2i\beta'_1})$$
$$b_4^{on-subs} = e^{-i\beta'_1}t_{10}e^{-i\beta'_1}r''e^{-i\beta'_1}r'_{10}e^{-i(\beta'_1-\beta'_x)}r'' = b_2^{on-subs}(r''r'_{10}e^{-2i\beta'_1})$$
$$\vdots$$
$$\Rightarrow$$

$$b_{2n+1}^{on-subs} = b_1^{on-subs}(r''r'_{10}e^{-2i\beta'_1})^n \qquad 23$$

$$b_{2n+2}^{on-subs} = b_2^{on-subs}(r''r'_{10}e^{-2i\beta'_1})^n \qquad 24$$

where $t_{10}$, $r'_{10}$, $\beta'_1$ and $\beta'_x$ are the same as the case for freestanding WS$_2$.

For on-substrate WS$_2$, the total amount of scattering becomes

$$F_{sc}^{on-subs} = \sum_{n\geq 0}[t_{10}e^{-i\beta'_x} + t_{10}r''e^{-i(2\beta'_1-\beta'_x)}](r''r'_{10}e^{-2i\beta'_1})^n \qquad 25$$

$$F_{sc}^{on-subs} = t_{10}\frac{e^{-i\beta'_x}+r''e^{-i(2\beta'_1-\beta'_x)}}{1+r''r'_{10}e^{-2i\beta'_1}} \text{ where } r'_{01} = -r'_{10}. \qquad 26$$

Similar to Equation 11, the intensity of WS$_2$ on-substrate is

$$I_{WS_2}^{on-subs} = \int_0^{d_1} |F_{abs}^{on-subs} \cdot F_{sc}^{on-subs}|^2 dx. \qquad 27$$

The last step is to define the enhancement factor by taking a ratio of Equation 27 and Equation 11 [2,5]

$$\Gamma = \frac{I_{WS_2}^{on-subs}}{I_{WS_2}^{susp}}. \qquad 28$$

With the help of the enhancement factor expression stated above, we calculated the enhancement factor ($\Gamma$) for hBN thickness $d_2 = 40nm$:

$$\Gamma|_{d_2=40nm} = \frac{I_{WS_2}^{on-subs}}{I_{WS_2}^{susp}} = 7.32427392.$$

Also, we calculated the enhancement factor as a function of the hBN thickness in nm $(1 \leq d_2 \leq 55)$:

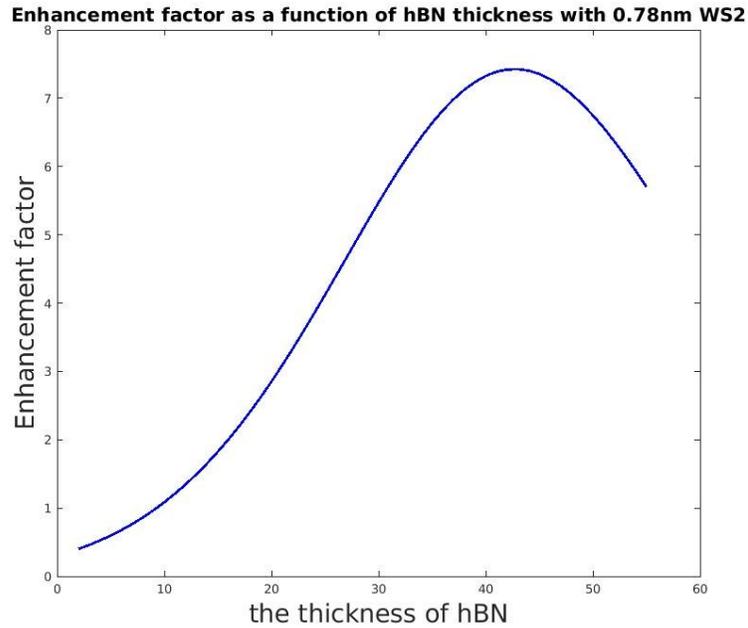

**Figure S8** Enhancement Factor as a function of hBN thickness[6].

3. **Gate dependent PL experiments**

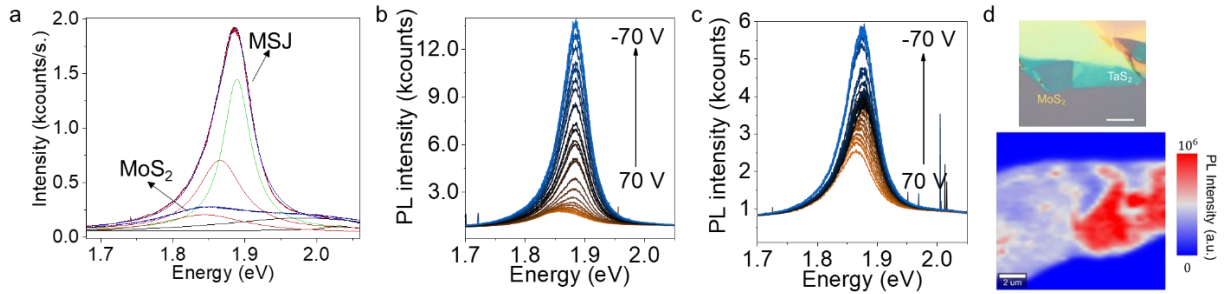

**Figure S9. Gate dependent PL measurements. a.** Gate dependent PL at 150 and 200 V for the sample presented in the main text. Fits to the PL spectra are also provided. **b.** PL spectra from 70 to -70 V gate voltage for bare MoS$_2$ and **c.** for TaS$_2$/MoS$_2$ on SiO$_2$/Si substrate. **d.** The optical microscope image and the PL intensity map is provided. The scale bar for the optical upper panel is 10 µm.

## 4. AFM Height Trace of Bilayer WS$_2$ Before and After Thermal Annealing

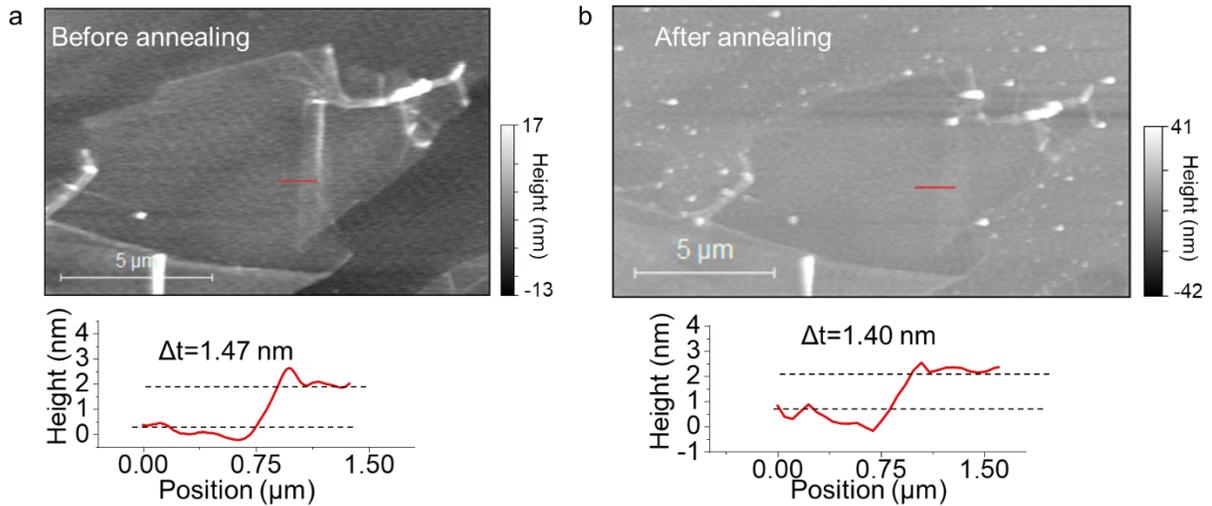

**Figure S10 | AFM Height Trace of Folded Part of monolayer WS$_2$. a.** AFM height trace from the folded part of the monolayer crystal shows 1.47 nm thickness over the monolayer and **b.** after annealing, the height trace from the bilayer region shows minimal change while the monolayer part height decreases by 0.50 nm.

## 5. XPS Measurements

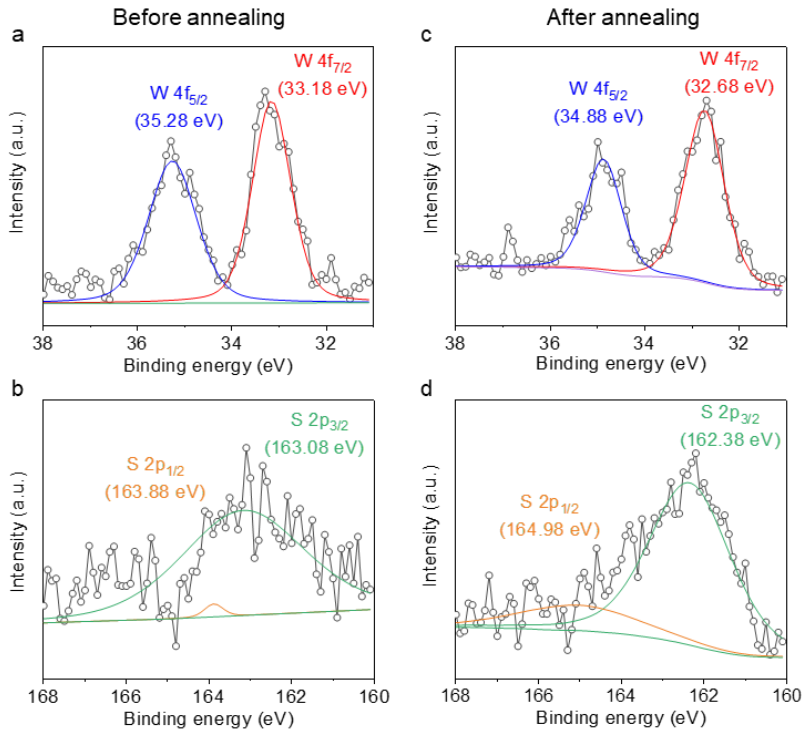

**Figure S11 Effect of annealing on the MSJ.** The XPS spectra of WS$_2$ multilayer **a.** W4f and **b.** S2p survey before annealing and **c.** W4f and **d.** S2p survey after annealing. The W peaks shift by 0.4 eV due to the charge transfer across the metal and the WS$_2$.

## 6. Temperature Dependent PL Measurements

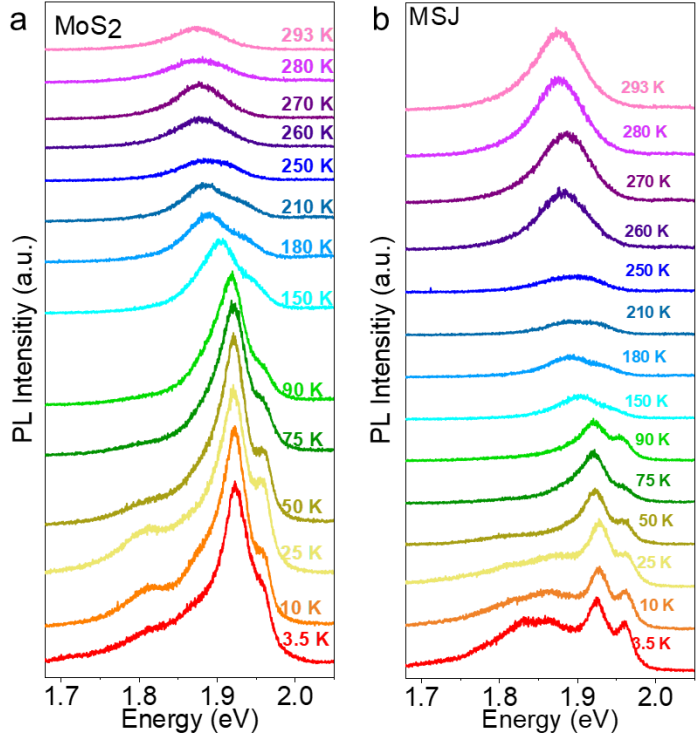

**Figure S12 | Temperature Dependent PL Spectra. a.** Temperature dependent PL spectra for monolayer MoS$_2$ and **b.** MoS$_2$/TaS$_2$ heterostructure. Below 250 K, enhancement disappears. As discussed in the main text, this can be attributed to the decrease in the van der Waals gap height as a result of the shrinkage of the hydrocarbon and water molecules between the layers. The PL spectra is shifted for viewing convenience. All spectra have the same absolute scale.

## 7. DFT Results

Before conducting the multi-layer investigation, we initially computed the electronic properties of the single-layer MoS$_2$, as illustrated in **Figure S13a**. As anticipated, the GW correction significantly affects the electronic energies, owing to the robust effective dielectric screening that alters the functional nature of the Coulomb interaction. The calculated band-gap value of 2.571 eV at the high-symmetry K point for this direct band-gap monolayer aligns well with prior calculations[7]. Additionally, the calculated quasiparticle peak energies, 1.96 and 2.10 eV are consistent with previous investigations[8] and the values obtained in our measurements.

Due to the substantial mismatch between the equilibrium in-plane lattice constants of TaS$_2$ and MoS$_2$, we initially examine the hetero-bilayer MoS$_2$-TaS$_2$ structure with MoS$_2$'s in-plane lattice parameters, while allowing all ions to relax along out-of-plane directions. In this context, we assume that our approach to synthesizing this hetero-bilayer structure without annealing does not create significant strain on the MoS$_2$ layer. Within the hetero-bilayer arrangement, the higher-lying valence states of the MoS$_2$ layer are located approximately 400 meV below the Fermi level, potentially enhancing exciton intensity, see **Figure S13b**. Surprisingly, the direct GW band gap corresponding to the MoS$_2$ layer reduces to 1.965 eV. Given this observation, one might anticipate a significant redshift in the direct exciton energy, yet we predict only minor meV redshifts in photoluminescence (PL) measurements. Generally, we would expect a noticeable alteration in the semiconductor layer due to the metallic screening around the Fermi level induced by the metallic substrate, whereas the observed reduction is significant. This prompt questioning the validity of the plasmon-pole approximation employed in our GW calculations. While this method typically performs well for semiconductors, its application to this hybrid structure (metal/semiconductor) may underestimate the band gap of the $MoS_2$ layer. Alternatively, the van der Waals distance between the $MoS_2$ and $TaS_2$ after direct transfer, without annealing, may be considerably greater than theoretical predictions.

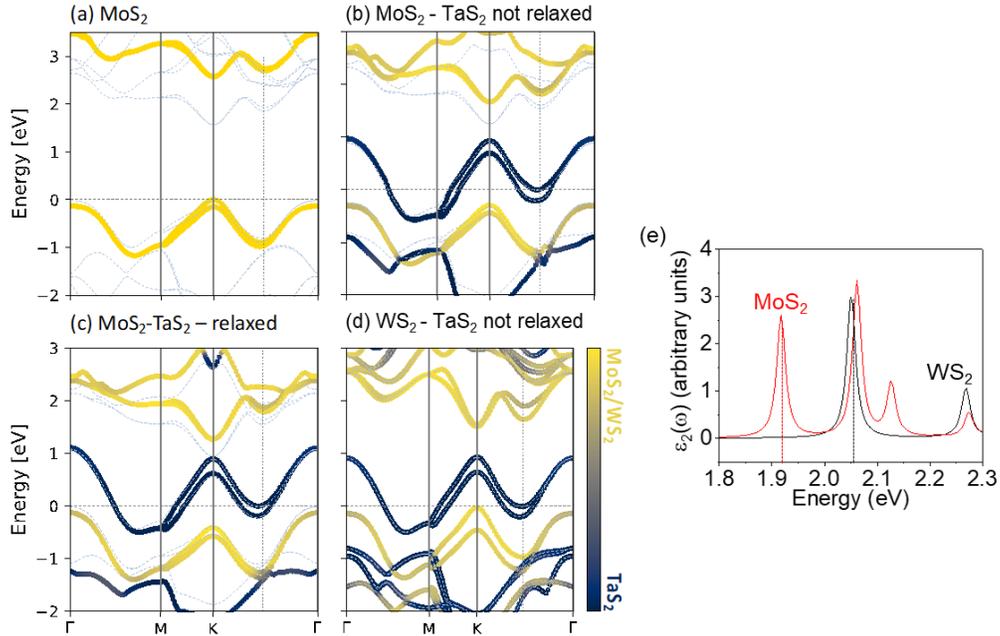

**Figure S13 | *ab initio* calculations.** The calculated band structures of **(a)** single layer $MoS_2$, **(b)** $MoS_2$-$TaS_2$ heterobilayer with single layer $MoS_2$ in-plane lattice constant, **(c)** $MoS_2$-$TaS_2$ heterobilayer with PBE relaxed lattice constants, and **(d)** $WS_2$-$TaS_2$ heterobilayer with single layer $WS_2$ in-plane lattice constant. The color bar represents the layer contributions to the corresponding electronic state. For $MoS_2$ structures, the dashed linesa and symbols represent the $G_0W_0$ and PBE calculations. For $WS_2$, only PBE results are presented. **(e)** The optical emission spectra for $MoS_2$ and $WS_2$ are given by the imaginary part of the excitonic macroscopic dielectric function $\epsilon_2$. Direct transition energy for excitons are 1.917 and 2.049 eV for $MoS_2$ and $WS_2$, respectively.

## 8. Photoluminescence Quantum Yield Measurements

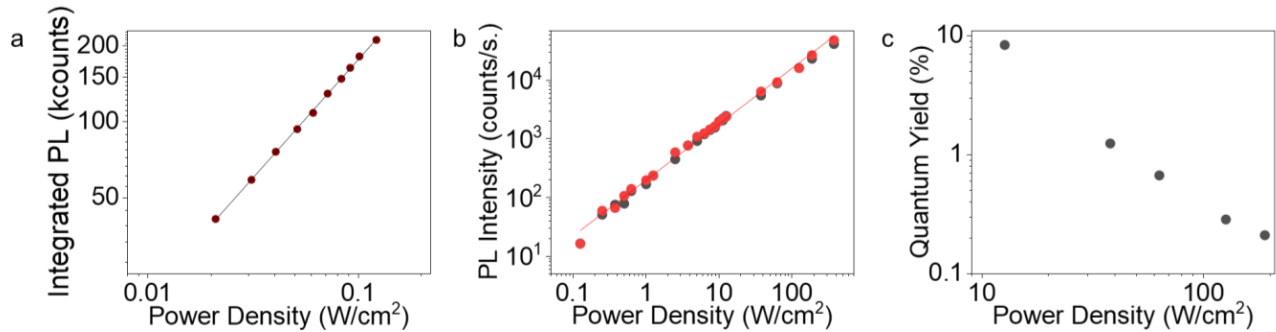

**Figure S14 | Quantum yield measurements. a.** Integrated PL counts vs power density is given for R6G embedded PMMA film on glass substrate. PL spectra is collected using an integrating sphere. Laser beam is focused to a $mm^2$ sized Gaussian spot. As a result, lower power densities are achieved that the tightly focused beam. **b.** PL intensity vs. power density measurements collected from a $WS_2$ monolayer on Au substrate. Red and grey dots represent PL intensity counts collected from different parts of the sample. **c.**

Quantum yield calculations performed for the sample provided in **b** using the calibration data from the R6G crystals and the calibration data. All PL spectra is collected with 532 nm excitation laser.